\documentclass[prl,twocolumn,showpacs,superscriptaddress]{revtex4-2}  
\usepackage{graphicx}
\usepackage{mathrsfs}
\usepackage{dcolumn}
\usepackage{bm}
\usepackage{amsmath}
\usepackage{amsfonts}
\usepackage{color}

\begin{document}

\title{Chiral Topological superconductivity in the OAI/SC/FMI heterostructure avoiding the subband problem}

\author{Jingnan Hu}
\thanks{These authors made equal contributions to this work.}
\affiliation{Wuhan National High Magnetic Field Center $\&$ School of Physics, Huazhong University of Science and Technology, Wuhan 430074, China}
\author{Fei Yu}
\thanks{These authors made equal contributions to this work.}
\affiliation{Wuhan National High Magnetic Field Center $\&$ School of Physics, Huazhong University of Science and Technology, Wuhan 430074, China}
\author{Aiyun Luo}
\affiliation{Wuhan National High Magnetic Field Center $\&$ School of Physics, Huazhong University of Science and Technology, Wuhan 430074, China}
\author{Jinyu Zou}
\affiliation{Wuhan National High Magnetic Field Center $\&$ School of Physics, Huazhong University of Science and Technology, Wuhan 430074, China}
\author{Xin Liu}
\affiliation{Wuhan National High Magnetic Field Center $\&$ School of Physics, Huazhong University of Science and Technology, Wuhan 430074, China}
\affiliation{Institute for Quantum Science and Engineering, Huazhong University of Science and Technology, Wuhan, 430074, China}
\affiliation{Wuhan Institute of Quantum Technology, Wuhan, 430074, China}
\author{Gang Xu}
\email[e-mail address: ]{gangxu@hust.edu.cn}
\affiliation{Wuhan National High Magnetic Field Center $\&$ School of Physics, Huazhong University of Science and Technology, Wuhan 430074, China}
\affiliation{Institute for Quantum Science and Engineering, Huazhong University of Science and Technology, Wuhan, 430074, China}
\affiliation{Wuhan Institute of Quantum Technology, Wuhan, 430074, China}

\begin{abstract}
Implementing topological superconductivity (TSC) and Majorana states (MSs) is one of the most significant and challenging tasks in both fundamental physics and topological quantum computations. In this work, taking the obstructed atomic insulator (OAI) Nb$_3$Br$_8$, s-wave superconductor (SC) NbSe$_2$ and ferromagnetic insulator (FMI) as example, we propose a new setup to realize the 2D chiral TSC and MSs in the OAI/SC/FMI heterostructure, which could avoid the subband problem effectively and has the advantage of huge Rashba spin-orbit coupling. As a result, the TSC phase can be stabilized in a wide region of chemical potential and Zeeman field, and four distinct TSC phases with superconducting Chern number $\small{\mathcal{N}= -1, -2, -3, 3}$ can be achieved. Moreover, a 2D BdG Hamiltonian based on the triangular lattice of obstructed Wannier charge centers, combined with the s-wave superconductivity paring and Zeeman field, is constructed to understand the whole topological phase diagram analytically. These results expand the application of OAIs and pave a new way to realize the TSC and MSs with unique advantages.
\end{abstract}

\maketitle

Topological superconductivity (TSC) has attracted intensive interest for its ability to host the Majorana states (MSs) and its implementation of topological quantum computations~\cite{majorana1937teoria,kitaev2001unpaired,alicea2012new,Sau2011controlling,kitaev2003fault,xu2014topological,zou2021new,zhang2019helical,zhang2019higher,zhang2021intrinsic,giwa2021fermi,zhang2013time,yang2014dirac,wang2018high,margalit2022chiral,kawakami2015evolution}. The initial system to realize TSC involves the p-wave pairing superconductor (SC)~\cite{nelson2004odd,Maeno1994Superconductivity,Hor2010Superconductivity,jiao2020chiral}. However, intrinsic p-wave SCs are very rare and MSs have not been clearly distinguished in these systems. In the past decade, many proposals are raised to induce TSC at the surface/interface of the topological materials~\cite{fu2008superconducting,hosur2011majorana,peng2019proximity,xu2016topological,pan2019lattice,liu2022prediction}, semiconductors with Rashba spin-orbit coupling (Rashba-SOC)~\cite{lutchyn2010majorana,oreg2010helical,stanescu2011majorana,sau2010generic,alicea2010majorana}, or magnetic atom chains~\cite{nadj2013proposal,kim2014helical,klinovaja2013topological,li2014topological,glazov2014exciton} by the proximity effect with the s-wave SC. Among them, one delicate proposal is to realize the TSC and MSs, including the Majorana zero modes and chiral MSs, in the 1D (nanowire)~\cite{lutchyn2010majorana,oreg2010helical,stanescu2011majorana} and 2D~\cite{sau2010generic,alicea2010majorana} semiconductor/s-wave SC heterostructures by modulating the Rashba-SOC, Zeeman field, and chemical potential elaborately to satisfy an odd number of subbands occupation. Moreover, it is proposed that the chiral MSs are easier to realize non-Abelian quantum gate operations, which could be 10$^3$ faster than the currently existing quantum computation schemes~\cite{lian2018topological,sau2010generic,alicea2010majorana}.

\begin{figure}[htb]
\centering
\includegraphics[width=0.5\textwidth]{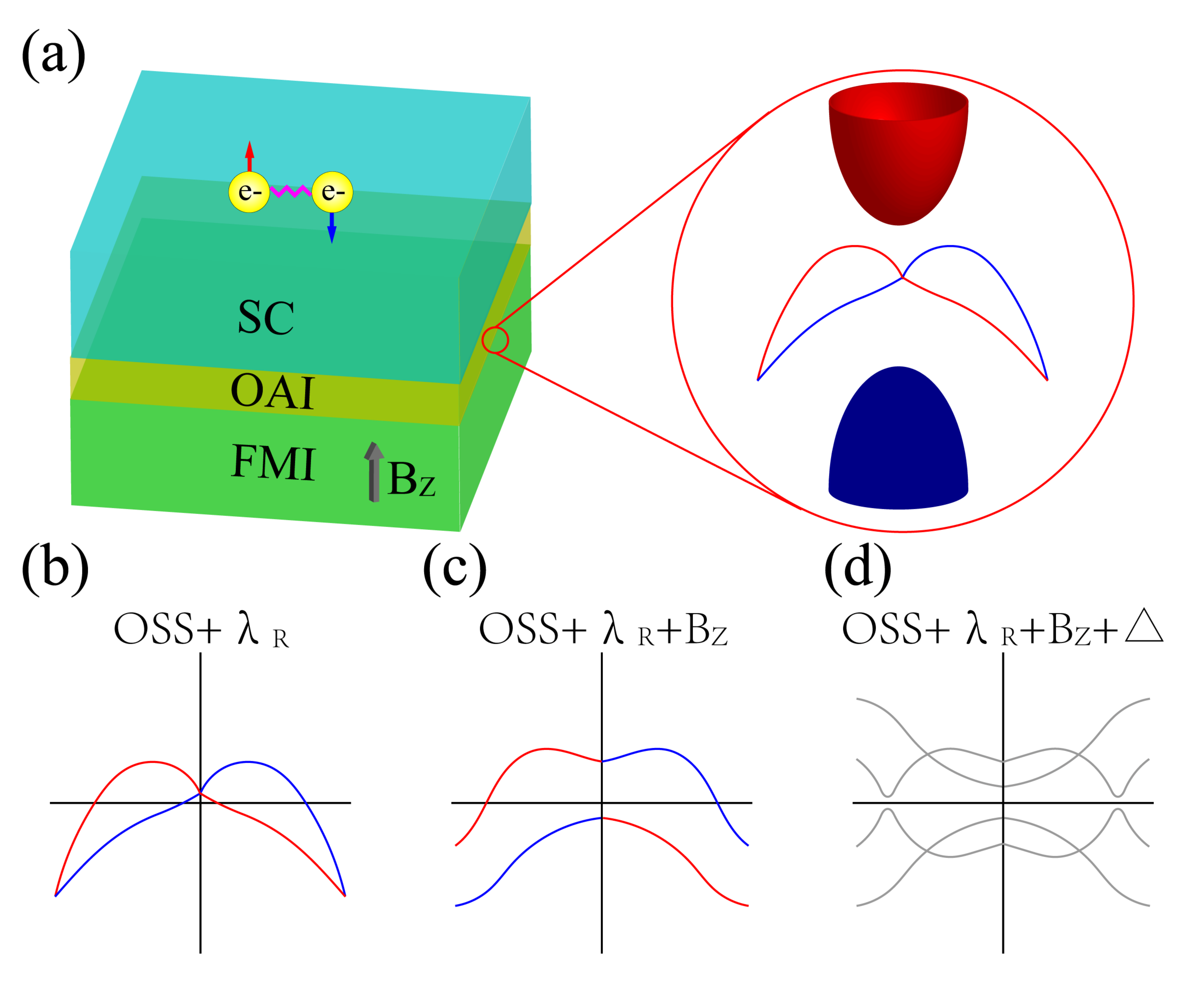}
\caption{(a) Schematic of the new chiral TSC setup based on SC/OAI/FMI heterostructure, where the bulk states of the OAI and the corresponding OSSs at the interface are illustrated in the red circle. (b) OSSs exhibit huge Rashba-SOC splitting. (c) The Kramers degeneracy of OSSs lifted by the out-of-plane Zeeman field $B_z$. (d) Illustration of the TSC spectrum realized on the OSSs incorporating with the Zeeman field $B_z$ and SC pairing $\Delta$.}
\end{figure}

Recently, many important advances have been made in the semiconductor/s-wave SC heterostructures~\cite{mourik2012signatures,deng2012anomalous,das2012zero,churchill2013superconductor,finck2013anomalous,vaitiekenas2021zero}. However, it also encounters many difficulties and challenges, mainly caused by the multiple subbands and rather small SOC~\cite{ford2012observation,vigneau2014magnetotransport,halpern2015room,fasth2007direct,liang2012strong,hernandez2010spin,iorio2018vectorial,scherubl2016electrical}. The multiple subbands not only reduce the energy window of odd number subbands occupation to reach the TSC phase~\cite{woods2020subband}, but also brings out many trivial low-energy Andreev bound states that mimic the Majorana conductance signals~\cite{bagrets2012class,liu2012zero,degottardi2013majorana,adagideli2014effects,woods2019zero,chen2019ubiquitous,rainis2013towards,pikulin2012zero}. Meanwhile, the small Rashba-SOC splitting would affect the stability of s-wave superconducting pairs in the presence of Zeeman field, thus make great limitation of experimental confirmation of MSs~\cite{weggemans2021tunable}.

In this work, by constructing the s-wave SC/obstructed atomic insulator (OAI)/ferromagnetic insulator (FMI) heterostructure as shown in Fig. 1(a), we propose a new TSC setup that could avoid the subband problem, in which the 2D TSC could be achieved by incorporating obstructed surface states (OSSs) with s-wave pairing and Zeeman field. Generally, OAIs are one kind of insulators that contain the obstructed Wannier charge centers (OWCCs), which are mismatched with the atomic positions but pinned at the Wyckoff positions between two atoms. As a result, metallic surface states, referred as OSSs, should exist inevitably on the OAI's surface that crosses the OWCCs, though it is topologically trivial insulator. Such OSSs are usually well separated from the bulk bands, and exhibit considerable Rashba-SOC splitting due to the spontaneous inversion symmetry breaking at the surface/interface, as schematically shown in Fig. 1(a) and (b). Therefore, OSSs are good analogs to the 2D electron gas at the interface of semiconductor/SC heterostructures~\cite{kalisky2012critical,lee2019transport}. The out-of-plane Zeeman field will lift the Kramers degeneracy at the time-reversal invariant momentum points, which can lead to an odd number of Fermi surfaces (FSs) as shown in Fig. 1(c). Since opposite spins are locked at the opposite momentum of OSSs, the s-wave pairing would be introduced into OSSs easily by the proximity effect, according to Fu-Kane argument~\cite{fu2008superconducting}. Consequently, it has a big chance to generate the TSC phase on such OSSs with superconducting pairing and Zeeman field as schematically shown in Fig. 1(d).

\begin{figure}
\centering
\includegraphics[width=0.5\textwidth]{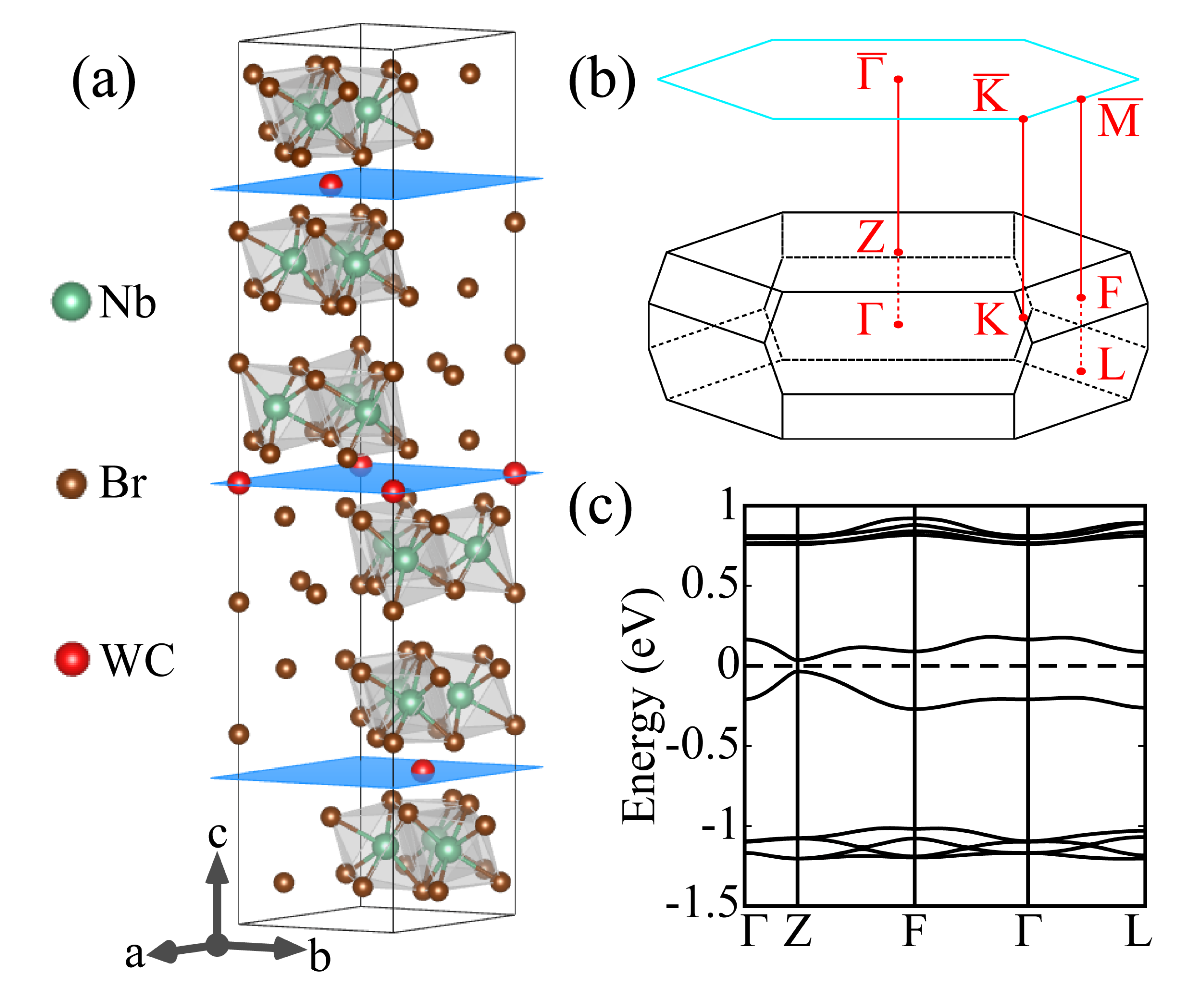}
\caption{(a) Crystal structure of Nb$_3$Br$_8$, where the red balls represent the positions of the OWCCs. The blue planes are the cleavage surfaces that cross the OWCCs. (b) High-symmetry points of the 3D BZ of the primitive cell Nb$_3$Br$_8$ and the projected 2D BZ on the $(1,1,1)$ surface. (c) Bulk band structures with SOC along the high-symmetry paths of Nb$_3$Br$_8$.}
\end{figure}

Here we choose Nb$_3$Br$_8$ as a concrete example to verify the feasibility and advantages of our proposal. As shown in Fig. 2(a), Nb$_3$Br$_8$ is a van der Waals layered material that crystallizes in the 166 space group, where Wyckoff position 6c and 18h are occupied by Nb and Br atoms respectively, and its corresponding Brillouin zone (BZ) is plotted in Fig. 2(b)~\cite{simon1966beta,meyer1994elektronische,habermehl2010triniobiumoctabromide}. As reported in Ref.~\cite{xu2021three}, Nb$_3$Br$_8$ is identified as an OAI with OWCCs located at empty Wyckoff position 3b as sketched by the red balls in Fig. 2(a). These OWCCs form the triangular lattice between the van der Waals layers as shown by the blue planes in Fig. 2(a). As a result, a natural cleavage surface of Nb$_3$Br$_8$ usually consists of such OWCCs, leading to the 2D metallic OSSs at the surface or interface.
Moreover, superconducting diode effects have been reported in the NbSe$_2$/Nb$_3$Br$_8$/NbSe$_2$ Josephson junctions recently~\cite{wu2022field}, which indicates the possibility to introduce the s-wave pairing into the OSSs at the interface of the experimentally synthesized Nb$_3$Br$_8$/NbSe$_2$ heterostructure.

By using the experimental crystal parameters~\cite{simon1966beta}, we perform the electronic structures calculations of Nb$_3$Br$_8$ based on the Vienna \emph{ab initio} simulation package~\cite{kresse1993ab,kresse1996efficiency} with 500~eV energy cutoff, $9\times 9\times 9$ k-point grids and the Perdew-Burke-Ernzerhof type exchange-correlation potential~\cite{perdew1996generalized}. The calculated band structures including SOC are plotted in Fig. 2(c), which gives rise to a 0.69~eV insulating gap, consistent with previous works very well~\cite{xu2021three}. By calculating the band representation at the maximal high-symmetry k-points, we confirm that Nb$_3$Br$_8$ is a topologically trivial insulator and an OAI. To confirm the OAI character of Nb$_3$Br$_8$, we build a 20-layers slab and calculate the slab bands by WannierTools~\cite{wu2018wanniertools} based on the Wannier functions of the Nb-d orbits constructed by WANNIER90~\cite{mostofi2008wannier90}. The slab bands are plotted in Fig. 3(a), in which the metallic OSSs originated from the OWCCs on the surface always present at the Fermi level ($E_F=0$), no matter SOC is considered or not. The OSS calculated without SOC is displayed in grey in Fig. 3(a), while the OSSs considering SOC are plotted by the color bar to represent spin $S_x=1/2$ (red) and $S_x=-1/2$ (blue) component, respectively. These results manifest that the OSS on the surface of Nb$_3$Br$_8$ has a strong Rashba-SOC splitting, which can be estimated roughly by the band splitting between the OSSs with and without SOC. By measuring the energy difference of the OSS bands maximum along the $\bar{\Gamma}-\bar{\mathrm{M}}$ path as shown in Fig. 3(a), the amplitude of Rashba-SOC splitting $\xi_{SO}$ is evaluated as about $6.7\mathrm{~meV}$, which is about one order larger than that in semiconductor InAs~\cite{fasth2007direct,liang2012strong,hernandez2010spin,iorio2018vectorial,scherubl2016electrical}.

Such strong Rashba-SOC will result in robust spin-momentum locking OSSs. We calculate the spin textures of the FSs at 4 meV above the Fermi level $E_F$ (green dotted line in Fig. 3(a)), as plotted in Fig. 3(b), which reveals that the spins at opposite momentum are oriented oppositely due to the strong Rashba-SOC effect. These results make it much easier to introduce the s-wave pairing into the OSSs through the proximity effect~\cite{weggemans2021tunable}. Generally, the out-of-plane Zeeman field $B_z$ will orient the spin to the $z$-direction and destroy the in-plane spin-momentum locking. However, the strong Rashba-SOC in Nb$_3$Br$_8$ will keep the spin-momentum locking OSSs to survive under the relatively large out-of-plane Zeeman field. In Fig. 3(c) and (d), we plot the OSSs on the bottom layer and the corresponding spin textures of the FSs with $B_z=10~\mathrm{meV}$ applied on the bottom layer, which unveil that OSSs still have sizable in-plane spin components. These results demonstrate that the superconducting proximity effect could mantain on the OSSs of Nb$_3$Br$_8$ even under the relatively large Zeeman field.

Now we go further to investigate the TSC properties of the SC/OAI/FMI heterostructure as shown in Fig. 1(a). The BdG Hamiltonian is constructed based on the 20-layers Nb$_3$Br$_8$ slab by adding the s-wave pairing $\Delta$ on all layers and applying the Zeeman field $B_z$ on the bottom layer. Fig. 4(a) shows the low energy spectrum of the BdG Hamiltonian with chemical potential $\mu=0$, Zeeman field $B_z=2.5~\mathrm{meV}$ and SC pairing amplitude $\Delta=1~\mathrm{meV}$ (pairing amplitude in NbSe$_2$). The superconducting spectrum in Fig. 4(a) is fully gapped. So the superconducting Chern number $\mathcal{N}$ can be well defined~\cite{sato2017topological}, which can be characterized by the winding number of the occupied BdG spectrum's Wilson loop~\cite{yu2011equivalent,soluyanov2011computing,gresch2017z2pack}. The corresponding Wilson loop evolution of all occupied states in Fig. 4(a) is calculated and plotted in Fig. 4(b), which winds 1 time over the whole range of $k_x$ obviously, confirming $\mathcal{N}=-1$ directly. These results manifest that chiral TSC can be achieved on the OSSs by incorporating with the s-wave pairing and tiny Zeeman field, and verify the feasibility of our new TSC setup proposed in Fig. 1(a).

\begin{figure}
\centering
\includegraphics[width=0.5\textwidth]{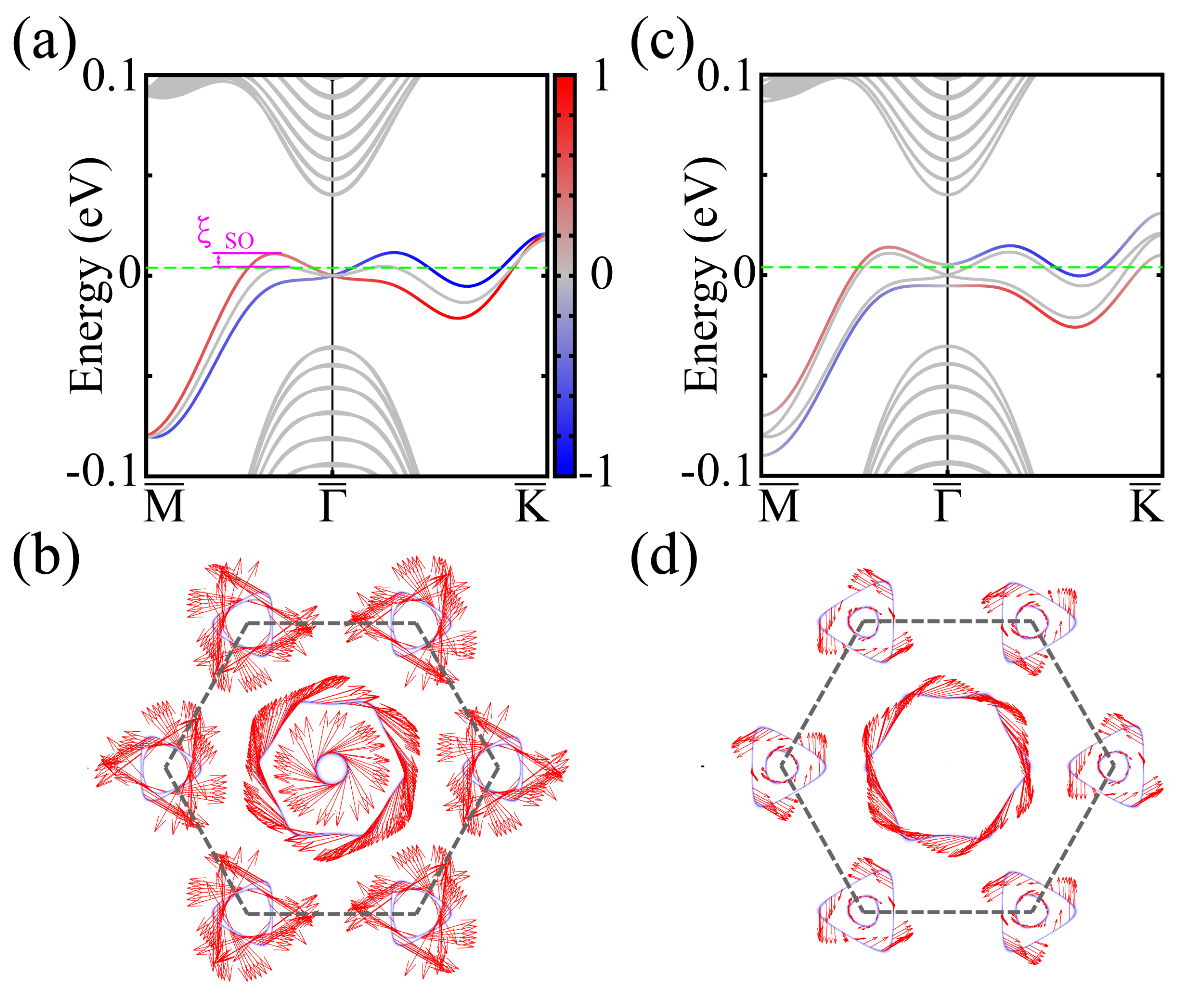}
\caption{(a) Band structures of $20$ layers Nb$_3$Br$_8$ slab at $B_z$ = 0, where OSS without SOC is plotted in grey and OSSs with SOC are plotted by the color bar to describe different spin $S_x$ components. $\xi_{SO}$ indicates the amplitude of Rashba-SOC splitting. (b) The spin textures on the FSs when the Fermi energy in (a) is set at 4 meV above $E_F$ (green dotted line). (c) Band structures of $20$ layers Nb$_3$Br$_8$ slab with $10\mathrm{~meV}$ out-of-plane Zeeman field applied on the bottom layer. The grey OSSs and color OSSs are from the top layer and bottom layer respectively. (d) Spin textures on the FSs when Fermi energy in (c) is set at 4 meV above $E_F$.}
\end{figure}

One can understand the TSC properties of the heterostructure through a 2D triangular lattice model analytically, since the OSSs are originated from the OWCCs that form the triangular lattice as sketched in Fig. 2(a). The 2D tight-binding model based on the triangular lattice with one orbital per site is constructed as the form of $\hat{H}_{SS}=\sum_{k,\sigma,\sigma^{'}}c_{k \sigma}^{\dagger}{[H_{SS}(k)]}_{\sigma {\sigma}^{'}}c_{k \sigma^{'}}$ ($\small{\sigma,\sigma^{'}=\uparrow,\downarrow}$) and  $\small{H_{SS}(k)= \varepsilon(k)+\sum_{j=x,\,y}E_{R}^{\left(j\right)}(k)\sigma_{j}}$ with
\begin{equation}
\begin{aligned}
 \varepsilon(k) &= \varepsilon_{1}(k)+\varepsilon_{2}(k)+\varepsilon_{3}(k)-\varepsilon_{0} \\
 \varepsilon_{1}(k) &= -2 t_{1}[\cos \left(2k_{x}L\right)+2 \cos \left(k_{x}L\right) \cos (\sqrt{3}k_{y}L)] \\
 \varepsilon_{2}(k) &= -2 t_{2}[\cos (2\sqrt{3}k_{y}L)+2 \cos \left(3k_{x}L\right) \cos (\sqrt{3}k_{y}L)] \\
 \varepsilon_{3}(k) &= -2 t_{3}[\cos \left(4k_{x}L\right)+2 \cos \left(2k_{x}L \right) \cos (2\sqrt{3}k_{y}L)] \\
 E_{R}^{x}(k)& =2 \lambda_{R}\sqrt{3}\cos \left(k_{x}L\right)\sin (\sqrt{3}k_{y}{L}) \\
 E_{R}^{y}(k)& =-2 \lambda_{R}[\sin \left(2k_{x}L \right)+\sin \left(k_{x}L\right) \cos (\sqrt{3}k_{y}L)]
\end{aligned}
\end{equation}
where $\sigma_{i}$ $(i=x,y,z)$ are the Pauli matrices in the spin space. $\varepsilon_{0}$ is the on-site energy that can be contracted into the chemical potential in the subsequent BdG Hamiltonian. ${t}_{1}$, ${t}_{2}$, ${t}_{3}$ are the hopping amplitudes of the nearest neighbor, next nearest neighbor, and third nearest neighbor. $L=a/2$ is length parameter described by the lattice constant $a$. $E_{R}^{x/y}(k)$ is the $x/y$ component of the Rashba-SOC term with $\lambda_{R}$ describing the strength of Rashba-SOC. We use such an effective model to fit the numerically calculated OSSs band structure of $\mathrm{Nb}_{3}\mathrm{Br}_{8}$ in Fig. 3(a) and obtain the fitted parameters as $t_{1}=-5~\mathrm{meV}$, $t_2=-5~\mathrm{meV}$, $t_3=7~\mathrm{meV}$, and $\lambda_{R}=2.5~\mathrm{meV}$.

Taking the Zeeman field $B_z$ and the SC pairing amplitude $\Delta$ into account, we can write out the total 2D effective Hamiltonian as $\small{\hat{H}=\hat{H}_{SS}+\hat{H}_{Z}+\hat{H}_{SC}}$ with $\hat{H}_{Z}=\sum_{k,\sigma,\sigma^{'}} c_{k \sigma}^{\dagger}{[B_z\sigma_{z}]}_{\sigma {\sigma}^{'}}c_{k \sigma^{'}}$ and $\hat{H}_{SC}=\sum_{k}\Delta(c_{k \uparrow}^{\dagger} c_{-k \downarrow}^{\dagger}+H.c.)$. Here we set $B_z>0$ without losing the generality. In the Nambu basis $\Psi_{k}=(c_{k \uparrow}, c_{k \downarrow}, c_{-k \uparrow}^{\dagger},c_{-k \downarrow}^{\dagger})^{T}$, this total Hamiltonian $\hat{H}= \small{\frac{1}{2}\sum_{k}\Psi_{k}^{\dagger} H_{BdG}(k) \Psi_{k}}$ can be rewritten as
\begin{equation}\label{eq:eq1}
	H_{BdG}=\tau_{z}\otimes(E_{T}\sigma_{0}+E_{R}^{y}\sigma_{y}+B_z\sigma_{z})+E_{R}^{x}\tau_{0} \otimes \sigma_{x}-\Delta \tau_{y} \otimes \sigma_{y}
\end{equation}
where $\tau_{i}$ ($\sigma_{i}$) and $\tau_{0}$ ($\sigma_{0}$) are the Pauli and identity matrices in the particle-hole (spin) space. $\small{E_{T}(k)=\varepsilon(k)-\mu}$ and $\mu$ is the chemical potential. By solving Eq. (2) analytically, it is straightforward to prove that the energy spectrum can only be closed at the high-symmetry points $\bar{\Gamma}$, $\bar{M}$, and $\bar{K}$ of the 2D BZ as shown in Fig. 2(b). Therefore, the topological phase transition can only be determined by the effective $k\cdot p$ model $\mathcal{H}_{\bar{\Gamma}}$, $\mathcal{H}_{\bar{M}}$ and $\mathcal{H}_{\bar{K}}$ by treating $\Delta$ as perturbation~\cite{Sato2010}. To do that, we first write the unperturbed Hamiltonian $\small{\hat{H}_{0}=\hat{H}_{SS}+\hat{H}_{Z}}$, and its spectrum in Nambu basis is expressed as
\begin{equation}\label{eq:spectrum}
  \pm\varepsilon_{\pm}(k)=\pm(\varepsilon(k)-\mu \pm \delta \varepsilon(k))
\end{equation}
with $\delta \varepsilon=\small{\sqrt{(E_{R}^{x})^{2}+\small{(E_{R}^{y})^{2}}+B_z^{2}}}$. $\small{\chi_{1}^\pm= (E_{R}^{x}-i E_{R}^{y},\pm\delta\varepsilon-B_z,0,0)^T/N}$ and $\small{\chi_{2}^\pm=(0,0,-E_{R}^{x}-iE_{R}^{y},\pm\delta \varepsilon-B_z)^T/N}$ are the eigenvectors corresponding to $\varepsilon_{\pm}$ and $-\varepsilon_{\pm}$ with the normalization factor $\small{N=\sqrt{2\delta\varepsilon(\delta\varepsilon\mp B_z)}}$. In the following, we will construct the effective $k\cdot p$ model near the high-symmetry points with the perturbation of $\Delta$, to demonstrate the system's TSC properties.

We first construct the effective model near $\small{\bar{\Gamma}}$ point to explain the TSC phase realized in Fig. 4(a) and 4(b). By setting the Zeeman field as a fixed value $B_{\bar{\Gamma}}$, the topological phase transition would be just controlled by the chemical potential $\mu$~\cite{qi2010chiral}. The gap closing condition at $\small{\bar{\Gamma}}$  point $\varepsilon_{\pm}(\bar{\Gamma})=0$  divides the whole chemical potential into two distinct topological regions, light blue area satisfying  $\varepsilon(\bar{\Gamma})+ B_{\bar{\Gamma}}>\mu>\varepsilon(\bar{\Gamma})- B_{\bar{\Gamma}}$  as shown in Fig. 4(c) and the region otherwise. While the latter chemical potential region gives rise to the topologically trivial SC, TSC phase can be realized in the light blue area~\cite{qi2010chiral}, in which the two lowest eigenvectors $\chi_{1,2}^-$ are used to derive the minimal Hamiltonian as $[\mathcal{H}_{\bar{\Gamma}}]_{\alpha \beta}=\langle \chi_{\alpha}^- | (\hat{H}_{0}+\hat{H}_{SC})|\chi_{\beta}^- \rangle$ ($\alpha,\beta=1,2$). Finally, we obtain
\begin{equation}\label{kp}
	\mathcal{H}_{\bar{\Gamma}}(k)= -\tilde{\Delta}(k_{x}\sigma_{x}-k_{y}\sigma_{y})+(\frac{1}{2m_{\bar{\Gamma}}}k^2-\mu_{\bar{\Gamma}})\sigma_{z}
\end{equation}
with $k^2=k_x^2+k_y^2$.
Eq.~(\ref{kp}) is a canonical 2D chiral p-wave superconductivity model with
$\tilde{\Delta}=3a\Delta\lambda_{R}/B_z$, $\mu_{\bar{\Gamma}}=\mu-\varepsilon(\bar{\Gamma})+B_{\bar{\Gamma}}$ and
$m_{\bar{\Gamma}}=1/3a^2(t_1+3t_2+4t_3)>0$,
which gives rise to a $\small{\mathcal{N}=-1}$ TSC phase according to the definition of the SC Chern number $\mathcal{N}=-(sgn[m_{\bar{\Gamma}}]+sgn[\mu_{\bar{\Gamma}}])$~\cite{shen2012topological}.
Thus, Eq.~(\ref{kp}) well explains the TSC physics when the chemical potential is set near the $\bar{\Gamma}$ point ($\mu=0$), i.e., the TSC phase realized in Fig. 4(a) and (b).

\begin{figure}
	\centering
	\includegraphics[width=0.5\textwidth]{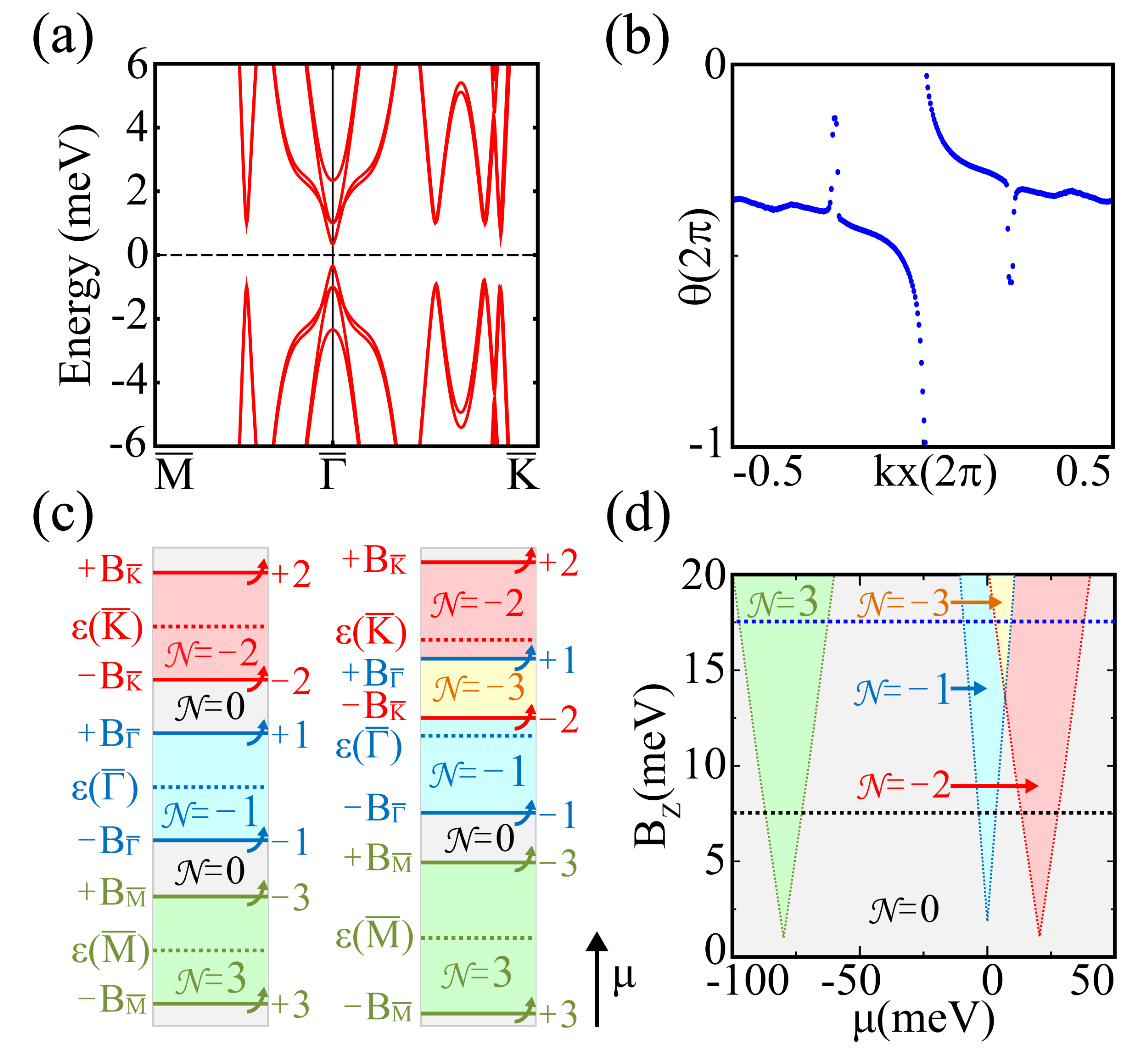}
	\caption{(a) BdG spectrum of Nb$_3$Br$_8$ slab with $\mu=0$, $B_z=2.5~\mathrm{meV}$ and $\bigtriangleup=1~\mathrm{meV}$. (b) The corresponding Wilson loop evolution of all occupied states in (a). (c)Illustration of the TSC phase evolution with $\mu$ increasing, where $B_{\bar{\mathrm{M}},\bar{\Gamma},\bar{\mathrm{K}}}$ mean the applied Zeeman field at the $\bar{\mathrm{M}},\bar{\Gamma},\bar{\mathrm{K}}$
points. The left and right panels correspond to the TSC evolution at weak and strong $B_z$ situations, as illustrated by the black and blue dash line in (d) respectively. (d) The calculated whole TSC phase diagram of the heterostructure in the space of $B_z$ and $\mu$ by using $\Delta = 1$ meV.}
\end{figure}

Similar discussion can be made at $\small{\bar{\mathrm{M}}}$ and $\small{\bar{\mathrm{K}}}$ points.
The gap closing condition near $\small{\bar{\mathrm{M}}}$ point $\varepsilon_{\pm}(\bar{\mathrm{M}})=0$ will give the TSC phase area enclosed by $\mu=\varepsilon(\bar{\mathrm{M}})\pm B_{\bar{\mathrm{M}}}$ (light green area in Fig. 4(c)),  where $B_{\bar{\mathrm{M}}}$ is the fixed Zeeman field at $\bar{\mathrm{M}}$ point. By using two lowest eigenvectors $\chi_{1,2}^-$, the minimal model in the light green area can be deduced as $\mathcal{H}_{\bar{\mathrm{M}}}(k)= -\tilde{\Delta}(k_{x}\sigma_{x}+k_{y}\sigma_{y})+(\frac{1}{2m^x_{\bar{\mathrm{M}}}}k^2+\frac{1}{2m^y_{\bar{\mathrm{M}}}}k^2-\mu_{\bar{\mathrm{M}}})\sigma_{z}$,
with $\mu_{\bar{\mathrm{M}}}=\mu-\varepsilon(\bar{\mathrm{M}})+B_{\bar{\mathrm{M}}}$, $m^x_{\bar{\mathrm{M}}}=1/a^2(t_1-9t_2+12t_3)>0$ and $m^y_{\bar{\mathrm{M}}}=1/3a^2(-t_1+t_2+4t_3)>0$, which is obversely a chiral p-wave SC with $\small{\mathcal{N}=1}$.
Moreover, since the Berry curvature near three $\small{\bar{\mathrm{M}}}$ points accumulate with each other~\cite{Kezilebieke2020}, $\small{\mathcal{N}=3}$ topological phase is realized in this area. At $\small{\bar{\mathrm{K}}}$ points, the lowest eigenvectors become
$\chi_{1,2}^+$ to determine topological phase in the red area enclosed by $\mu=\varepsilon(\bar{\mathrm{K}})\pm B_{\bar{\mathrm{K}}}$ in Fig. 4(c) with $B_{\bar{\mathrm{K}}}$ as the fixed Zeeman field at $\bar{\mathrm{K}}$ point.
The $k\cdot p$ Hamiltonian is thus given as $\mathcal{H}_{\bar{\mathrm{K}}}(k)= -\tilde{\Delta}/2(k_{x}\sigma_{x}+k_{y}\sigma_{y})+(\frac{1}{2m_{\bar{\mathrm{K}}}}k^2-\mu_{\bar{\mathrm{K}}})\sigma_{z}$, with
$\mu_{\bar{\mathrm{K}}}=\mu-\varepsilon(\bar{\mathrm{K}})-B_{\bar{\mathrm{K}}}$ and $m_{\bar{\mathrm{K}}}=2/3a^2(-t_1+6t_2-4t_3)<0$, which gives rise to SC Chern number $-1$ at each $\bar{\mathrm{K}}$ point. Since there are two equivalent $\small{\bar{\mathrm{K}}}$ points,
 $\small{\mathcal{N}=-2}$ TSC phase is realized in the red area finally.

According to the energy arrangement of the $\bar{\Gamma}$, $\small{\bar{\mathrm{M}}}$ and $\small{\bar{\mathrm{K}}}$ shown in Fig. 3(a) and (c), the TSC phase should evolute with $\mu$ increasing as illustrated in the left panel of Fig. 4(c), where three TSC phases determined by $\mathcal{H}_{\bar{\Gamma}}$, $\mathcal{H}_{\bar{M}}$ and $\mathcal{H}_{\bar{K}}$ are well separated from each other, corresponding to the weak $B_z$ situation. Thus we can tune the chemical potential $\mu$ to get three different TSC phases. Moreover, if the applied $B_z$ is strong enough so that the areas overlap, an extra TSC phase with $\small{\mathcal{N}=-3}$ will emerge as suggested by the yellow area in the right panel of Fig. 4(c).

In Fig. 4(d), we further numerically calculated the whole TSC phase diagram of the heterostructure as function of $\mu$ and $B_z$ by using the same parameters in the calculations of Fig. 4(a) and 4(b),
which agree with our model analysis very well except that the TSC phase can only be obtained when the Zeeman field exceeds a threshold value.This is because the effective Zeeman gap should surpass the energy of $\Delta$ at least to achieve a nontrivial SC Chern number~\cite{qi2010chiral}. Finally, it is also worth noticing that some in-gap segmented FSs will appear in the $\bar{\Gamma}-\bar{\mathrm{K}}$ path due to the Ising type SOC at large Zeeman field $B_z$, as reported in Ref. \cite{Zhu2021}.

In conclusion, we have proposed a new setup to realize the 2D chiral TSC in the SC/OAI/FMI heterostructure. The OSSs of OAI possess strong Rashba-SOC and are well separated from the bulk bands. These advantages result in a wide chemical potential range to create the clean TSC phase that could efficiently avert the subband problem and the trivial Andreev bound states. We consider the OAI $\mathrm{Nb}_{3}\mathrm{Br}_{8}$ and s-wave SC $\mathrm{NbSe}_2$ as concrete example to verify the feasibility and advantages of our new proposal, whose heterostructure have been synthesized successfully. By performing the $\emph{ab initio}$ calculations and the slab BdG Hamiltonian calculations, we have numerically simulated the SC spectrum at the interface of the heterostructure, and then obtained the Wilson loop evolution of all occupied BdG spectrum, which obviously confirm that TSC characterized by $\small{\mathcal{N}=-1}$ can be achieved at the natural carrier filling interface ($\mu=0$) with the accessible Zeeman field ($\small{B_z=2.5~\mathrm{meV}}$). Moreover, we also construct a 2D effective BdG model based on the triangular lattice of OWCCs. By treating the SC pairing $\Delta$ as perturbation, three minimal $k\cdot p$ models $\mathcal{H}_{\bar{\mathrm{\Gamma}}}, \mathcal{H}_{\bar{\mathrm{M}}}$ and $\mathcal{H}_{\bar{\mathrm{K}}}$ are further derived, and the whole TSC phase diagram in the space of $\mu$ and $B_z$ is figured out analytically, which demonstrates that four distinct TSC phases with SC Chern number $\small{\mathcal{N}= -1, -2, -3, 3}$ can be achieved in the whole $\mu$ and $B_z$ space. These results provide OAIs as a new platform to design the chiral TSC and MSs, which will stimulate more interests from both the TSC and OAI studies, especially the studies on $\mathrm{Nb}_{3}\mathrm{Br}_{8}$/$\mathrm{NbSe}_2$ heterostructure.

This work is supported by the National Key Research and Development Program of China (2018YFA0307000), and the National Natural Science Foundation of China (12274154,11874022).

\bibliography{refs}

\end{document}